# A Smartphone Controlled Handheld Microfluidic Liquid Handling System

Baichen Li,[1] Lin Li,[2,3] Allan Guan,[1] Quan Dong,[1] Kangcheng Ruan,[2,3] Ronggui Hu[2,3] and Zhenyu Li[*1]

## Abstract

Microfluidics and lab-on-a-chip technologies have made it possible to manipulate small volume liquids with unprecedented resolution, automation and integration. However, most current microfluidic systems still rely on bulky off-chip infrastructures such as compressed pressure sources, syringe pumps and computers to achieve complex liquid manipulation functions. Here, we present a handheld automated microfluidic liquid handling system controlled by a smartphone, which is enabled by combining elastomeric on-chip valves and a compact pneumatic system. As a demonstration, we show that the system can automatically perform all the liquid handling steps of a bead-based sandwich immunoassay on a multi-layer PDMS chip without any human intervention. The footprint of the system is $6 \times 10.5 \times 16.5$ cm, and the total weight is 829g including battery. Powered by a 12.8V 1500mAh Li battery, the system consumed 2.2W on average during the immunoassay and lasted for 8.7 hrs. This handheld microfluidic liquid handling platform is generally applicable to many biochemical and cell-based assays requiring complex liquid manipulation and sample preparation steps such as FISH, PCR, flow cytometry and nucleic acid sequencing. In particular, the integration of this technology with read-out biosensors may help enable the realization of the long-sought Tricorder-like handheld in-vitro diagnostic (IVD) systems.

## Introduction

An important goal of lab-on-a-chip research is to miniaturize conventional biological and chemical instruments into chip formats[1-3]. One common feature of such instruments is that they all require or rely on the handling of liquid samples such as blood, urine, and liquid reagents etc. Therefore, it is essential for lab-on-a-chip systems to have built-in liquid handling capabilities comparable to conventional manual micropipetting or robotic liquid handling in order to achieve truly automated sample-to-answer operations. Indeed, in the past two decades, there have been extensive research efforts to miniaturize liquid handling components on chip, such as MEMS valves and pumps[4,5], elastomeric on-chip valves and pumps[6], and droplet manipulation systems[7]. Compared with conventional liquid handling systems, such miniaturized technologies have enabled the manipulation of small volume liquids with unprecedented resolution (picoliter or less)[8], automation[9] and parallelism[10]. However, due to various fabrication, integration and reliability challenges, few *handheld* self-contained microfluidic systems capable of complex liquid handling exist in the market today except for capillary-driven microfluidics[11]. Most lab-on-a-chip systems still rely on bulky off-chip infrastructures such as compressed pressure sources, syringe pumps and computers to achieve their liquid manipulation functions. Recently Sia et al. developed a handheld instrument which can actuate on-chip elastomeric microvalves using solenoid-containing actuation units[12]. Another promising technique is digital microfluidics in which microdroplets are manipulated by electrowetting[7]. Additionally, Braille display devices have been used to build portable microfluidic cell culture systems[13]. In this paper, we present a smartphone-controlled handheld microfluidic liquid handling system by combining elastomeric on-chip valves and a handheld pneumatic system. The handheld pneumatic system provides on-board multiple pressure generation and control by using a miniature DC diaphragm pump, pressure-storage reservoirs, and small solenoid valves. This system is applicable to both single-layer pressure- driven microfluidics and multi-layer elastomeric microfluidics[6,14-16]. By *elastomeric microfluidics*, we mean microfluidic systems with on-chip valves based on the mechanical deformations of elastomeric membranes or structures, such as multi-layer PDMS microfluidics[6,15], glass/PDMS/glass devices[16] or other hybrid devices[14].

In a typical elastomeric microfluidic system, at least two different pressure sources are needed: one for actuating on-chip valves, which typically require a pressure level higher than 10 psi; and the other for driving reagents into microfluidic channels (for typical microfluidic channel dimensions, e.g. 10μm high, 100μm wide, 1 to 5 psi is sufficient). Traditionally, this is achieved by using two pressure regulators connected to a compressed gas tank[6]. However, the sizes and nature of these components make them unsuitable for building a handheld system. Although it is possible to use two miniaturized diaphragm pumps to build such a system, the significant fluctuations of the output pressure of a diaphragm pump limit its applications.

To address these challenges, we have developed a handheld microfluidic liquid handling system controlled by a smartphone (Fig.1), which can provide two different pressure sources and an array of 8 pneumatic control lines for operating elastomeric microfluidic chips. One pressure source is set to above 10 psi (max. 20 psi) to operate on-chip elastomeric valves; while the other can be set to any value below 5 psi to drive liquid flow, with a precision of ±0.05 psi. Eight independent pneumatic control lines are available to handle eight different reagents. The footprint of the resulting system is 6 × 10.5 × 16.5 cm, and the total weight is 829 g (including battery). The system can operate continuously for 8.7 hours while running an immunoassay liquid handling protocol when powered by a 12.8 V, 1500 mAh Li battery. This technology can serve as a general purpose handheld small volume liquid handling platform for many biochemical and cell-based assays such as fluorescence in-situ hybridization (FISH), PCR, flow cytometry and nucleic acid sequencing. The integration of this system with biosensors may help realize the long-sought dream of handheld multi-analyte in-vitro diagnostic (IVD) systems, i.e. Medical Tricorders[17].

**Experimental**

The overall handheld system consists of three subsystems: (1) a pneumatic pressure generation and control subsystem (Pneumatic subsystem); (2) an electronic printed circuit board (PCB) with 2 microcontrollers, a Bluetooth communication module, pressure sensors and power device drivers (Electronic subsystem); and (3) an elastomeric microfluidic chip (Microfluidic chip). The system can be controlled by a Bluetooth enabled Android smartphone (Galaxy S III). We describe each subsystem in details in the subsequent sections.

**Pneumatic subsystem**

The pneumatic subsystem is designed to generate two compressed air pressure sources at different levels (P1: >10 psi; P2: 0 to 5 psi) for operating elastomeric microfluidics. Two pressure reservoirs, labeled as Reservoir 1 and Reservoir 2 (Fig. 2), are used to store compressed air. A miniature DC diaphragm pump is used to pump air into Reservoir 1 to generate the primary pressure source for actuating on-chip elastomeric valves[6,14-16]. A secondary pressure source, stored in Reservoir 2, is derived from Reservoir 1 and stabilized by a feedback control system with a precision of ±0.05 psi for driving liquid reagents through microfluidic channels. The system can be easily extended to have multiple secondary pressure sources of different pressures if required.

Each pressure reservoir is made of four segments of 1/8'' ID Tygon tubing connected with a four-way barbed cross connector, leaving four open ports. Each open port of a reservoir is connected to a functional part of the pneumatic subsystem (such as a diaphragm pump, a solenoid valve or a pressure sensor, as shown in Fig. 2 and described in more detail below) via a barbed connector. The volume of each reservoir is determined by the total length of tubing used. In this work, the volumes of Reservoir 1 and 2 are 6.2 mL and 16.2 mL, respectively.

The four open ports of Reservoir 1 are connected to the following components, respectively:

(1) A miniature DC diaphragm pump (Parker, H004C-11), used to generate the primary pressure source for the system. A check valve in-between is used to prevent air leakage when the pump is off.
(2) Barometric Sensor 1, to monitor the pressure level in Reservoir 1.
(3) The normally open (N.O.) port of a solenoid valve manifold with eight channels (Pneumadyne, MSV10-8), each common port of this manifold is connected to an on-chip elastomeric valve.
(4) The common port of a solenoid valve (Valve 1, Pneumadyne, S10MM-30-12-3) with base (Pneumadyne, MSV10-1), to generate the secondary pressure source (P2, between 0 and P1, typically 0-5 psi). A plastic needle valve restrictor (Poweraire, F-2822-41-B85-K) is placed before the common port of Valve 1 to limit the air flow into Reservoir 2.

The four open ports of Reservoir 2 are connected to the components given below, respectively:

(1) The normally closed (N.C.) port of Valve 1 described above. Once the pressure in Reservoir 2 drops to below the lower bound of the target pressure range, Valve 1 will be opened to increase the Reservoir 2 pressure gradually. Another plastic needle valve restrictor (Poweraire, F-2822-41-B85-K) is used to fine tune the flow rate of air to reduce excessive pressure overshoot in Reservoir 2. The N.O. port of Valve 1 is completely sealed with PDMS to mimic a 2-way N.C. valve.
(2) The N.C. port of a solenoid valve (Valve 2) to release pressure in Reservoir 2. The common port of Valve 2 is sealed with PDMS, resulting in a small internal air volume (~250 μL). By toggling the state of Valve 2, its internal volume is connected to either Reservoir 2 or atmosphere. When Reservoir 2 is connected to Valve 2's internal volume, the pressure in Reservoir 2 drops by a small amount due to volume expansion. Immediately following this, the pressure in Valve 2's internal volume is released by connecting it to atmosphere. By choosing the appropriate Reservoir 2 volume, we can control the pressure release precisely to achieve the desired pressure resolution. Design details can be found in the ESI (S2-4).
(3) Barometric Sensor 2, to monitor the pressure level in Reservoir 2.
(4) The N.C. port of a solenoid valve (Valve 3), to drive reagents into a microfluidic chip. The common port of Valve 3 is divided into multiple lines, each connected to a reagent container (a modified micro-centrifuge tube). Once it is actuated, the pressure stored in Reservoir 2 is applied to every reagent container in the system.

**Microcontroller-based electronic subsystem**

We used two ATMega328p 8-bit microcontrollers in our PCB design. The first microcontroller is responsible for four tasks: (1) receive commands from a smartphone via Bluetooth 2.1 (Bluetooth Mate Silver, Sparkfun, WRL-12576), (2) pass commands to the second microcontroller, (3) request real-time barometric sensor data from the second microcontroller and send it to the smartphone at a frequency of 50 Hz, and (4) control the status of 6 load switch circuits for pump and solenoid valve operation. The second microcontroller also has four functions: (1) control the status of the other 12 load switch circuits (making the system capable of operating 18 electromechanical components (valves or pumps)), (2) monitor the barometric sensor data with the built-in 10-bit ADCs, (3) automatically adjust pressure levels in the two pressure reservoirs, and (4) send barometric sensor data to the first microcontroller on request. Communication between these two microcontrollers is realized with the Wire Library provided by Arduino, which enables bi-directional communication via only two wires using the I$^2$C protocol.

Two barometric sensors (Measurement Specialties, 1240-015D-3L) are used to acquire real-time pressure data in the pneumatic subsystem. Electrical sensor output data are amplified by two instrumentation amplifiers (Texas Instruments, INA114). We also used two general-purpose rail-to-rail operation amplifiers (Texas Instruments, OPA2171) to shift the signal level to between 0 and 5 V before connecting to the 10-bit ADCs of the second microcontroller. Each barometric sensor is calibrated with a commercial grade sanitary pressure gauge (Ashcroft, 1035). The calibration curve is saved in the microcontroller's Flash memory. Detailed calibration procedure and results can be found in the ESI (S4).

Our PCB circuit is powered by a 12.8 V $LiFePO_4$ battery pack with a capacity of 1500 mAh. This battery is directly used as the power supply to an array of 18 load switch circuits to power solenoid valves and the diaphragm pump. Each output of the load switch circuit is wired to a screw terminal block (Sparkfun, PRT-08084), into which the power wires of the valves and pump are mounted. The battery is also regulated by a 5V regulator to power the PCB (microcontrollers, Bluetooth module, Barometric Sensors & OpAmps).

**Pressure stabilization in Reservoir 1 and 2**

The pressure levels in the two reservoirs of the pneumatic subsystem are automatically adjusted by a feedback control algorithm programmed in the second microcontroller, with the barometric sensor as the feedback sensor, and the diaphragm pump and solenoid valves (Valve 1 & 2) for pressure control. A user only needs to set the upper and lower bounds of both pressure levels through the smartphone interface.

Reservoir 1, which contains higher pressure (>10 psi) to actuate on-chip valves and serves as the pressure source for Reservoir 2, is directly pressurized by the diaphragm pump. If the pressure level drops to below the chosen lower bound detected by Barometric sensor 1, the pump starts to run until the pressure is restored to the target upper bound.

The pressure in Reservoir 2 is used to drive liquid reagents through microfluidic channels, and the required pressure level is relatively low (normally below 5 psi for typical microfluidic channel dimensions of 10-100 μm). However, it is desirable to be able to precisely change the Reservoir 2 pressure in seconds. In order to obtain a stable Reservoir 2 pressure, instead of directly using the diaphragm pump as the pressure source, we used Reservoir 1 as a buffered pressure supply, connected to Reservoir 2 via a 2-way N.C. solenoid valve (Valve 1). As described above, another 3-way solenoid valve (Valve 2) is used to precisely release pressure in a stepwise fashion. If a pressure below the set lower bound is sensed, Valve 1 will be opened. Conversely, if the pressure exceeds the higher bound, Valve 2 starts to operate to decrease the pressure in Reservoir 2.

**Microfluidic device fabrication, reagent containers, and interfaces**

We fabricated a PDMS elastomeric microfluidic chip using multi-layer soft lithography[6] for bead-based sandwich florescence immunoassays. The on-chip valves were operated in a push-down configuration (flow layer on the bottom). We first fabricated the master molds using standard UV photolithography with AZ50XT positive photoresist. The control layer was made by pouring a PDMS mixture (RTV615 A:B at a 5:1 ratio) onto the mold and baking it for 1 hour at 65°C. For the flow layer, a PDMS mixture at a 20:1 ratio was spin-coated onto the flow layer mold at 3000 rpm for 1 minutes and baked at 65°C for 30 minutes. After curing of these two layers, we peeled off the control layer from the mold, aligned and placed it on top of the flow layer, followed by a 2 hour bake. Then the device was peeled off the flow layer mold and bonded to a glass slide by air plasma treatment. The design of the microfluidic device is given in the *Results and discussion* section below.

The reagent containers were modified 1.5mL disposable microcentrifuge tubes (Ted Pella MC-6600, polypropylene). The lid of every tube was punched by a 21 gauge needle to form two holes. Two 21 gauge needles of different lengths were inserted into the holes. The longer needle was immersed in the liquid reagent, while the shorter one was above the liquid surface for pressurization. By storing the reagents in the tubes, and applying air pressure through the shorter needle, the reagents were pushed out from the longer needle. Interfaces between the pneumatic subsystem, the reagent containers, and the microfluidic devices were made of Tygon tubings and 21 gauge needles. Details can be found in the ESI (S6).

**Simulated bead-based fluorescence immunoassay liquid handling**

As a demonstration, we used the system to perform all the liquid handling steps of a bead-based sandwich fluorescence immunoassay (FIA)[18]. We used three colored food dyes to simulate the reagents used in the immunoassay. The detailed 10-step FIA protocol (liquid handling sequences) is shown in Table 1 below. Briefly, the microfluidic channels are first blocked by a blocking buffer to reduce nonspecific binding. Then capture antibody-coated polystyrene microbeads are loaded and immobilized by the microfluidic hydrodynamic traps[19-21]. Microbeads of different sizes can be immobilized at different locations for multiplexed detections. Then after a washing step, sample solution is loaded into the reaction chamber and allowed to incubate. During the incubation, the liquid flow is controlled to move back and forth to facilitate mixing and target binding. The sample loading and incubation step is repeated 10 times. After another washing step, fluorescently labelled detection antibodies will be loaded and allowed for incubation. Finally excess detection antibodies will be washed away and the microbeads are then ready for fluorescence read-out.

**Table 1** Simulated Immunoassay Liquid Handling Protocol

1. Fill the device with blocking buffer (Green, 5s).
2. Incubate (5min).
3. Load washing buffer (Clear, 5min).
4. Load beads (Red, 10s)
5. Load washing buffer (Clear, 5min).
6. Load sample (Red, 5s) and incubate (1min).
   Repeat step 6 ten times.
7. Load washing buffer (Clear, 5min).
8. Load detection antibody (Green, 5s) and incubate (1min). Repeat step 8 ten times.
9. Load washing buffer (Clear, 5min).
10. Ready for detection

**Android smartphone app**

An Android (version 4.2) app with a graphical user interface (GUI) was written to allow user control of the system, and the display and analysis of collected data. We designed a straight-forward application protocol (details can be found in the ESI (S7)), and implemented it in both the smartphone app and ATMega328 microcontrollers. This application protocol has three basic functions: (1) set the status of each solenoid valve, (2) set the target pressure range of each reservoir, and (3) request barometric sensor readings from the microcontroller. The protocol of the simulated immunoassay is programmed in this Android application, using these three fundamental functions, to achieve liquid manipulation and monitor the pressure levels of the reservoirs.

## Results and discussion

We first characterized the pneumatic performance of the system, including achievable pressure range, pressure stability, response time, and leakage rate. Reservoir 1 was tested under no loading conditions, with a target pressure range of 13-14 psi. The maximum achievable pressure was ~20 psi limited by the diaphragm pump. Starting from 0 psi, it took about 0.3 second to rise to 14 psi (Fig. 3A). The pressure dropped by about 1 psi every minute, and the pump automatically started to restore the pressure in Reservoir 1. This level of pressure leakage, although preventable by better sealing, does not affect the actuation of on-chip valves, nor the pressure level in Reservoir 2 due to the feedback pressure stabilization mechanism described above.

Reservoir 2 was tested and characterized with a simple one-channel pressure driven microfluidic chip as the load. Colored food dye was pushed into the microfluidic channel (500 × 20 µm, W × H) continuously. We tested four integer pressure levels below 4 psi (1, 2, 3, 4 psi), with a tolerance of ±0.05 psi. The measured results are shown in Fig. 3B. It took about 10 seconds for the pressure in Reservoir 2 to rise to 4 psi, which is much longer compared with that of Reservoir 1. This is because we used two needle valve restrictors to limit the flow rate between Reservoir 1 and Reservoir 2 to improve stability. During the 10-minute test, the pressure was completely controlled in the target pressure ranges. At the end of the tests, we set the target pressure to 0 psi to examine the performance of pressure release. The results were comparable to theoretical calculations from ideal gas law.

During the tests, we found that the volume of a reservoir, especially Reservoir 2, had a significant influence on the performance. Whenever the system opens a valve, the effective volume of a reservoir will increase, which leads to a pressure drop inside the reservoir. When reagents are driven into a microfluidic device, the effective volume of Reservoir 2 is also increased (Fig. 3B). Increasing the volume of a reservoir will make it less sensitive to volume changes. However, this will increase the size of the system, which is undesirable for a handheld system. Moreover, increasing the volume of Reservoir 2 makes it respond slower when changing the target pressure settings. We found the parameters listed above gave sufficient performance for our immunoassay applications, but there is still room for improvements by optimizing the air flow rate and volumes of reservoirs.

Improving the sealing of the pneumatic system will also improve the overall performance of the system, in terms of pressure and power. In this work, no special effort was made to ensure that the system was completely airtight, as the pressure performance was sufficient for our applications. However, it is desirable and feasible to improve the sealing in the future with better pneumatic connections and components, so that the system can be more stable and consume less power.

Next, we tested the operation of on-chip valves under a bright field optical microscope. We validated that the on-chip valves could be fully closed without leakage by driving colored food dye into the flow layer at P2 = 2 psi, and toggling the state of the controlling solenoid valve multiple times. The on-chip valve was completely closed when P1 was set to 13-14 psi (Fig. 4). We also tested other PDMS devices with different valve geometries and configurations (both push-up and push-down), and found that it is possible to operate properly designed push-up valves with even lower actuation pressure down to 5 psi as reported previously[22].

Finally, we demonstrated a simulated immunoassay liquid handling protocol on a PDMS device automated by the handheld instrument and controlled by a Galaxy SIII smartphone. The design layout of the PDMS device is shown in Fig. 5A. The device has five inlet ports, two reaction chambers (sample and control), and one outlet port. Each inlet port is supplied with one of the following reagents: blocking buffer (e.g. PBS + 0.1% Tween-20 + 0.5% BSA), washing buffer (e.g. PBS + 0.1% Tween-20), capture antibody-coated polystyrene microbeads, sample, and fluorescently labelled detection antibodies. Inside the reaction chambers, arrays of hydrodynamic traps with openings of different sizes are used to immobilize individual capture antibody coated microbeads at predefined locations[18-20]. Sequential traps of different sizes (larger sizes upstream) can be used to trap different sized microbeads for multiplexed detection. In addition, such bead trapping based assays are naturally reconfigurable by changing the capture-molecule coated bead species without modifying the microfluidic chips. Once the microbeads are immobilized, subsequent mixing, washing and incubation steps can be easily performed on them under well controlled flow conditions. During the incubation steps, the liquid flow can be controlled to move back and forth by pressurizing port M (Fig. 5A) and the output port alternatively to facilitate mixing and target binding. For every test, all reagents except for the sample solution are loaded into both reaction chambers, following the sequence of a typical sandwich florescence immunoassay protocol. Sample solution will be loaded only into the sample chamber, leaving the other chamber as a control. All wastes is discharged from the outlet port, and collected by a disposable microcentrifuge tube. We designed a 10-step simulated fluorescence immunoassay protocol (Table 1), which took about 50 minutes to complete. This protocol was programmed in the Android application, and could be easily modified by the user. Two different colored food dyes and water were used to simulate various reagents. Screenshots of various liquids flowing inside the microfluidic device are shown in Fig. 5B. A video showing the system operating in real time is available in the ESI.

Power consumption is a critical performance metric of a handheld system. We used a digital multimeter (Keithley, Model 2000) to record the voltage and current changes at a sampling rate of 10 Hz while running the simulated immunoassay protocol. Power consumption was obtained by multiplying the voltage with the current (Fig. 6). The average power consumption during the simulated immunoassay was 2.2 W. As the capacity of the battery used was 19.2 Whr, the system could last for 8.7 hours on one full charge.

The average power consumption of 2.2W can be further reduced by 0.65W by replacing the normally closed (N.C.) solenoid valve (Valve 3) at the output of Reservoir 2 with a normally open (N.O.) one. Additionally, if we replace the solenoid valves with latching solenoid valves or use a 'Spike and Hold' circuit[23], the power consumption can be further reduced to less than 1 W. We also found that when our system was idle (no solenoid valves are operating), it still consumed 0.98W of power, which clearly can be reduced by redesigning the circuit with low power microcontrollers and electronic components. The system operation time can also be extended by using a higher capacity battery. For example, four 2600 mAh cell phone Li batteries (3.7V) can theoretically power the device for 25 hours on a full charge, thus permitting full day operation.

The current system footprint is $6 \times 10.5 \times 16.5$ cm (H $\times$ W $\times$ L), and the total weight is 829 g (including battery). By improving the design and choosing more compact components, we believe it is feasible to reduce the weight to below 1 pound (454g), and reduce the size by a half in the future. With better designed microfluidic devices, we can also decrease the pressure required for actuating on-chip valves, and consequently, further reduce the power consumption, size and weight of the whole system.

## Conclusion

In summary, we have demonstrated a smartphone-operated fully automated handheld pneumatic liquid handling system for elastomeric microfluidics. In addition to traditional multi-layer PDMS microfluidics, this system is also applicable to other types of elastomeric microfluidics such as devices similar to GE's Biacore[TM] SPR chips[14], glass/PDMS/glass devices[16] and other hybrid devices[24]. We believe this general purpose handheld liquid handling system is an enabling technology, which can find broad applications in point-of-care medical diagnostics, environmental testing, food safety inspection, biohazard detection, and biological research. In particular, the integration of this technology with read-out biosensors may one day help enable the realization of the long-sought Tricorder-like handheld IVD systems.

## Acknowledgements

This work was supported by the faculty start-up fund provided by the School of Engineering and Applied Science at The George Washington University.

## Notes and references

[1] *Nanophotonics and Microfluidics Laboratory, Department of Electrical and Computer Engineering, The George Washington University, Washington, DC 20052, USA. Tel: +1 (202) 994 4272; E-mail: zhenyu@gwu.edu*

[2]*State Key Laboratory of Molecular Biology*，[3]*Cancer Research Center, Shanghai Xu-Hui Central Hospital, Institute of Biochemistry and Cell Biology, Shanghai Institutes for Biological Sciences, 320 Yue-yang Road, Shanghai 200031, China.*

† Electronic Supplementary Information (ESI) available: Two videos showing the handheld system and the microfluidic chip in operation and a document describing the implementation details of the subsystems.

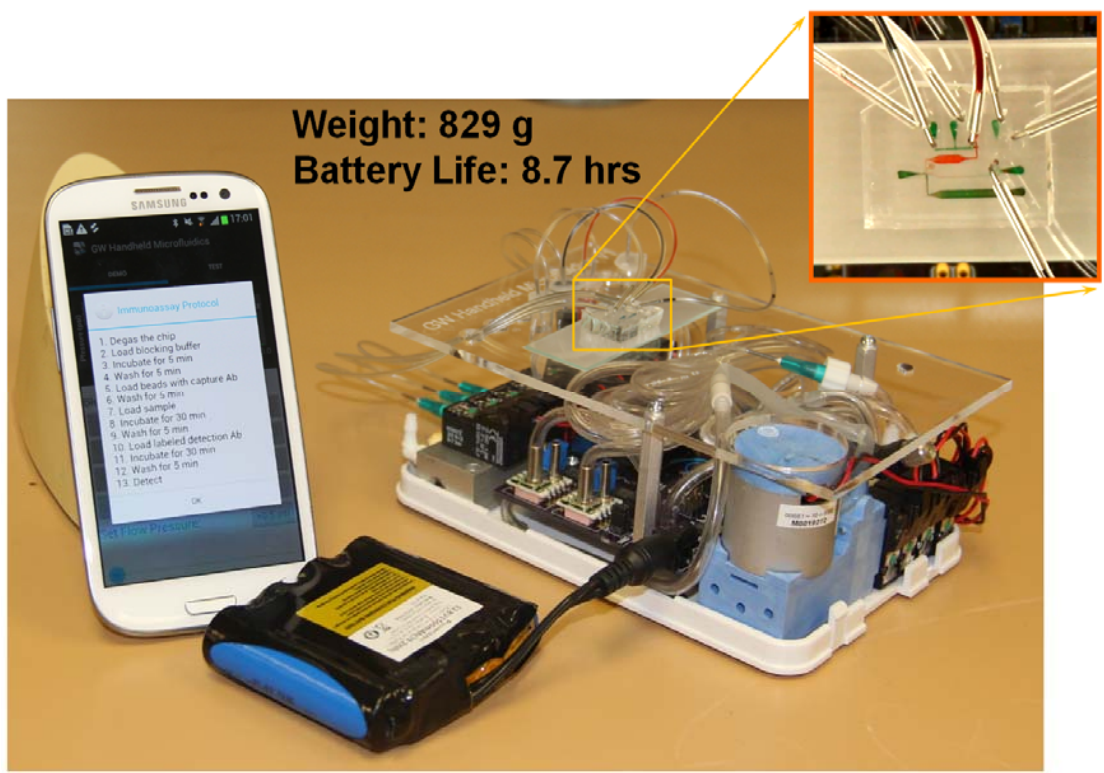

**Fig. 1** Picture of the smartphone-controlled handheld microfluidic liquid handling system. The footprint of the instrument is 6 × 10.5 × 16.5cm. Powered by a 12.8V 1500mAh Li battery, the instrument consumes 2.2W on average for a typical sandwich immunoassay and lasts for 8.7 hrs. Inset: a multi-layer PDMS device with on-chip elastomeric valves.

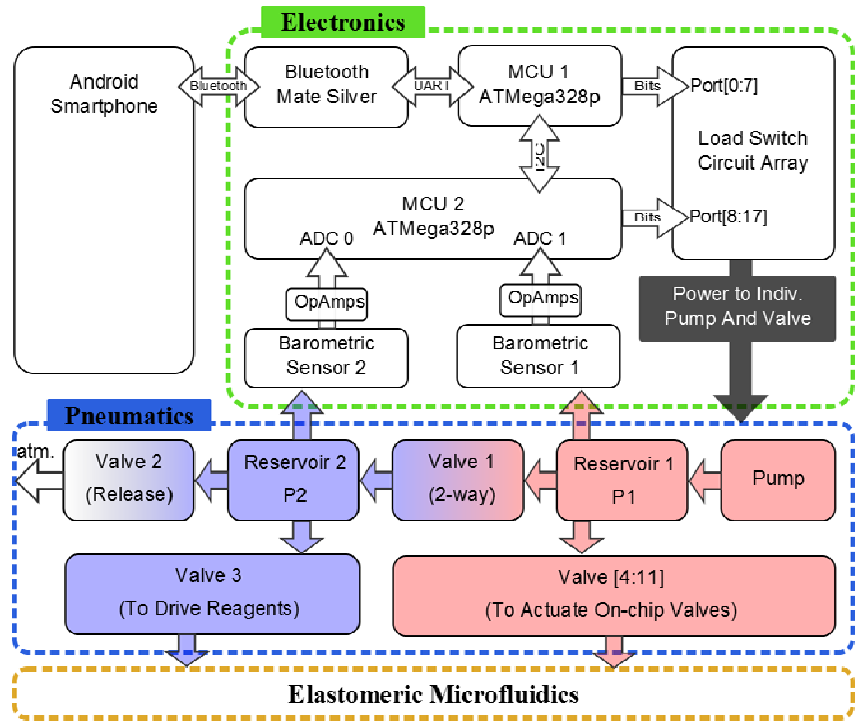

**Fig. 2** Block diagram of the system. The system consists of an Android smartphone, an electronic PCB with microcontrollers and Bluetooth module, and a pneumatic system capable of generating two different pressure output. Valve 1 is a two-way normally-closed solenoid valve and Valves 2 to 11 are three-way solenoid valves.

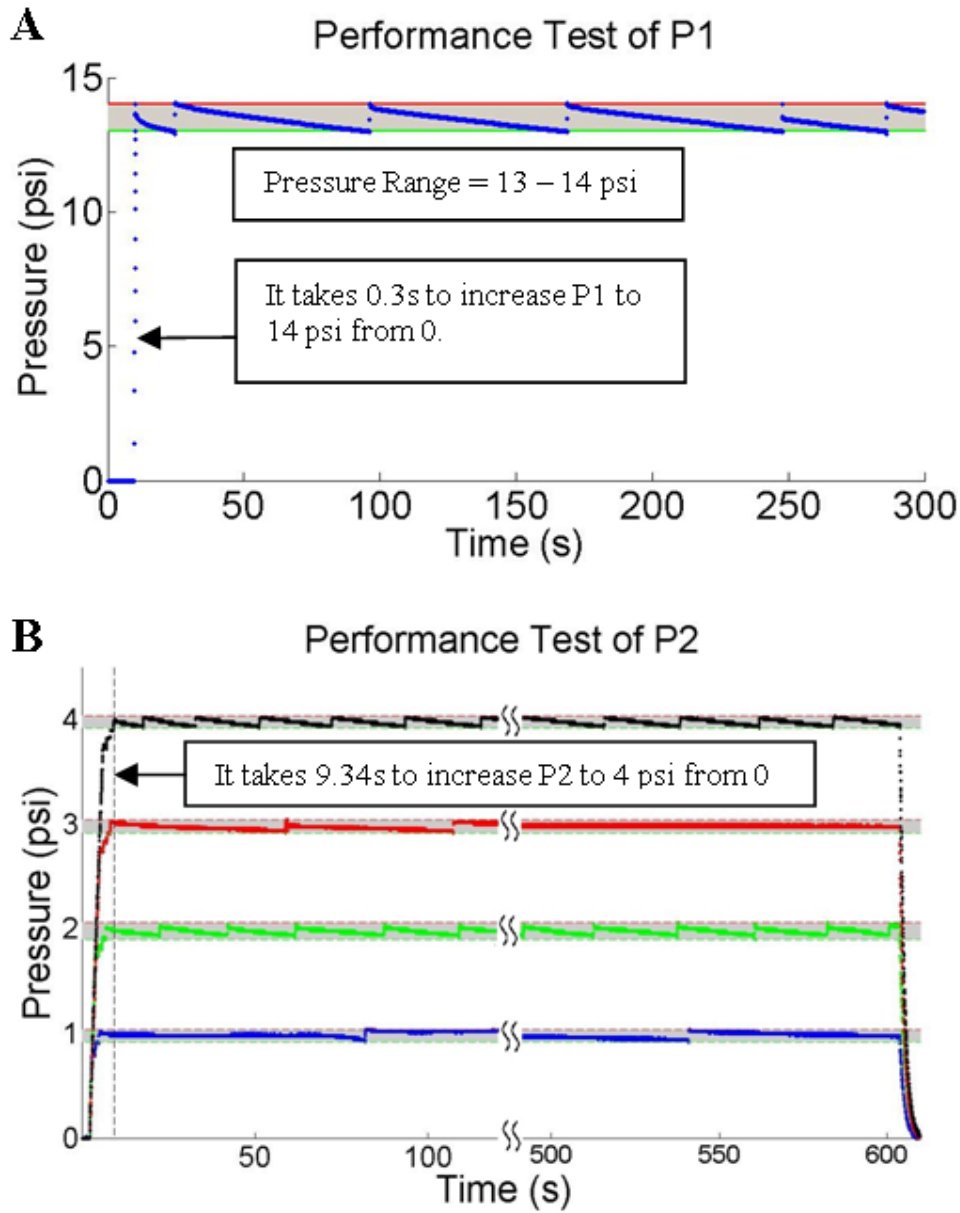

**Fig. 3** Pneumatic performance tests. (A) Performance test of P1. The target pressure range of P1 was set to 13-14 psi. The plot shows the reading from Barometric Sensor 1 over 300 s. (B) Performance test of P2. With P1 set to 13-14 psi, P2 was tested under four different target pressure levels: 1, 2, 3, and 4 psi (four colored curves), with a tolerance of ±0.05 psi. Meanwhile, one on-chip valve was opened to allow a reagent (colored food dye) to be driven into the microfluidic. Every test lasted for 10 minutes. The target pressure was set to 0 psi on completion of each test.

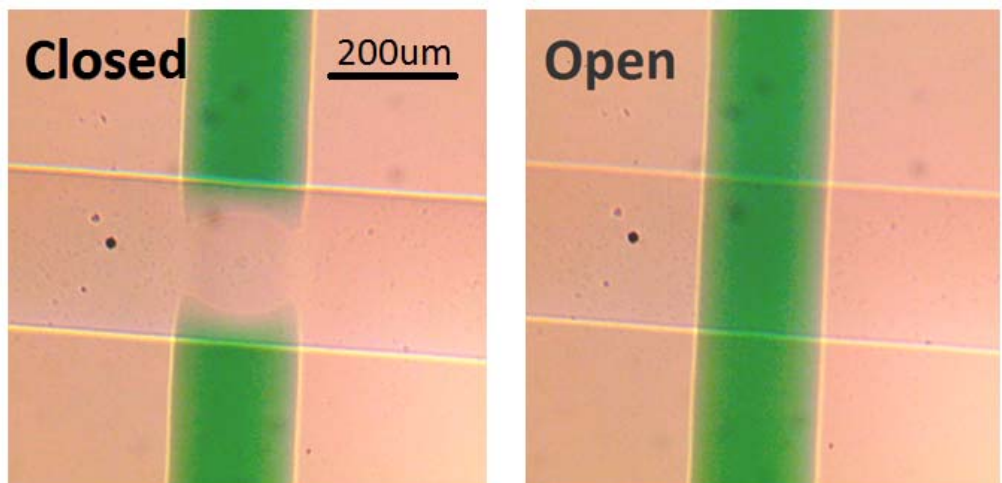

**Fig. 4** Demonstration of operation of an on-chip elastomeric valve under an optical microscope. Colored food dye was flowed through the microfluidic channel and the state of the on-chip valve was toggled. P1 was set to 13-14 psi, and P2 was set to 2 psi.

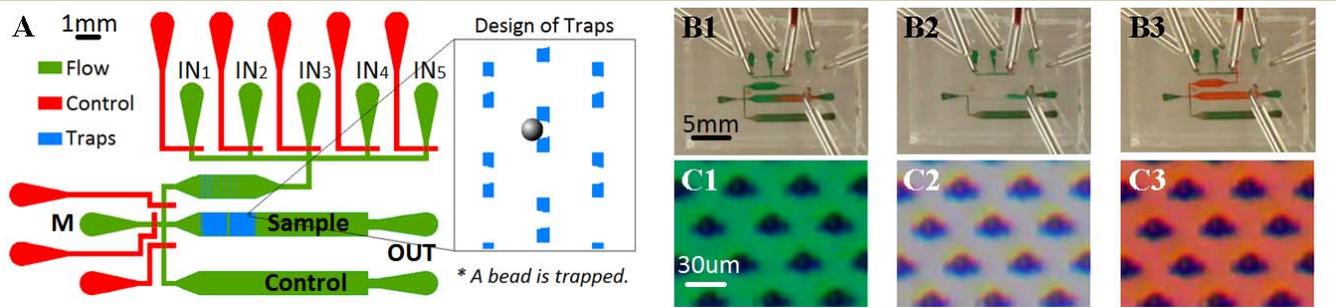

**Fig. 5** Two-layer PDMS microfluidic device for bead-based immunoassays and demonstration of immunoassay liquid handling. (A) Design layout of the microfluidic device for bead-based immunoassays. Inset: Magnified view of the hydrodynamic trap arrays with one trapped microbead. (B1-B3) Washing and incubating the immobilized microbeads with different reagents. Two colored dyes and water were used to simulate different reagents in an immunoassay. (C1-C3) Optical micrographs of a array of microbeads in the traps. Microbeads of 15μm diameter are loaded before the experiment. Optical micrographs showing the trapped microbeads being washed (or incubated) with different colored liquids. Two videos showing the operations of the system and the microfluidic traps are available in the ESI.

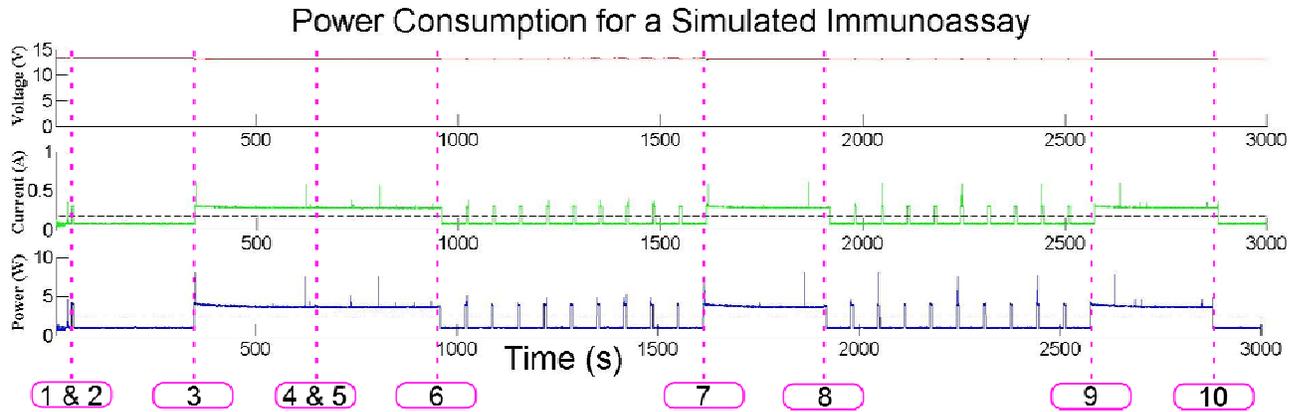

**Fig. 6** Power consumption for running the simulated immunoassay protocol (Table 1). A digital multimeter (DMM) was connected to the system to measure the voltage (*V*) and current (*I*) of the batter at a frequency of 10Hz. Power consumption was then calculated as *P* = *VI*. The immunoassay protocol took about 50 minutes to complete. Horizontal dashed lines (black) indicate the average values of the measured parameters. Vertical dashed lines (marked with numbers in a box) indicate the starting points of the various immunoassay steps. Refer to Table 1 for detailed descriptions of each step.